\documentclass[preprint]{aastex}

\begin{document}

\title{Detection of Near-IR CO Absorption Bands in R Coronae Borealis Stars}

\author{Emily D. Tenenbaum\altaffilmark{1,2}, Geoffrey C. 
Clayton\altaffilmark{2,3}, Martin 
Asplund\altaffilmark{4}, C.W. Engelbracht\altaffilmark{5}, Karl D. Gordon\altaffilmark{5}, M. M. Hanson\altaffilmark{6}, Richard J. Rudy\altaffilmark{7}, David K.ÊLynch\altaffilmark{7} , S. Mazuk\altaffilmark{7}, Catherine C. Venturini\altaffilmark{7}, and R. C. Puetter\altaffilmark{8} }

\altaffiltext{1}{Department of Chemistry, Pomona College, 645 N. College 
Avenue, Seaver North, Claremont, CA 91711-6338; emilytenenbaum@hotmail.com}
\altaffiltext{2}{Maria Mitchell Observatory, 3 
Vestal Street, Nantucket , MA 02554}
\altaffiltext{3}{Department of Physics \& Astronomy, Louisiana State
University, Baton Rouge, LA 70803; gclayton@fenway.phys.lsu.edu}
\altaffiltext{4}{Research School of Astronomy and Astrophysics, Mount 
Stromlo Observatory, Cotter Road, Weston, ACT 2611, Australia; 
martin@mso.anu.edu.au}
\altaffiltext{5}{Steward Observatory, University of Arizona, Tucson, AZ
85721;  cengelbracht, kgordon@as.arizona.edu}
\altaffiltext{6}{Department of Physics, University of Cincinnati, Cincinnati, OH, 45221-0011; hanson@physics.uc.edu}
\altaffiltext{7}{The Aerospace Corporation, M2/266, P.O. Box 92957, Los Angeles, CA 90009; david.k.lynch, richard.j.rudy, steve.mazuk, catherine.c.venturini@aero.org}
\altaffiltext{8}{UC, San Diego, CASS 0424, 9500 Gilman Dr., La Jolla, CA 92093-0424; rpuetter@ucsd.edu}

\begin{abstract}
R Coronae Borealis (RCB) stars are hydrogen-deficient,  carbon-rich pulsating post-AGB stars
that experience massive irregular declines in brightness caused by circumstellar dust 
formation. The mechanism of dust formation around RCB stars is not well understood. It has 
been proposed that CO molecules play an important role in cooling the circumstellar gas so that dust may form.  We report on a survey for CO in a sample of RCB stars.  We obtained H- and K-band  spectra including the first and second overtone CO  bands for 
eight RCB stars, the RCB-like star, DY Per and the final-helium-flash star, FG Sge. The first and second overtone CO bands were detected in the cooler (T$_{eff}<$6000 K) RCB stars, Z Umi, ES Aql, SV Sge and DY Per. The bands are not present in the warmer (T$_{eff}>$6000 K) RCB stars, R CrB, RY Sgr, SU Tau, XX Cam. In addition, first overtone bands are seen in FG Sge, a final-helium-flash star that is in an RCB-like phase at present. 
Effective temperatures of the eight RCB stars 
range from 4000 to 7250 K. The observed photospheric CO absorption bands were compared 
to line-blanketed model spectra of RCB stars. As predicted by the models, the 
CO bands are strongest in the coolest RCB stars and not present in the warmest. 
No correlation was found 
between the presence or strength of the CO bands and dust formation activity in the stars.
\end{abstract}


\section{Introduction}
R Coronae Borealis (RCB) stars are a small group of hydrogen-deficient, 
carbon-rich stars that have supergiant sizes.  Since the discovery of the prototype, R CrB, 
in 1795, about 50 RCB stars have been identified in the Galaxy and the Magellanic Clouds (Clayton 1996; Alcock et al. 2001; Tisserand et al. 2004).  
The major  defining 
characteristic of RCB stars is their unusual variability - they undergo 
massive declines of up to 8 magnitudes at irregular intervals.  The stars 
will typically stay at minimum light for weeks or months, before recovering to 
maximum brightness.  Formation of carbon-rich dust clouds is widely accepted as 
the cause of declines (Loreta 1934; O'Keefe 1939).  It is thought that amorphous carbon 
dust condenses 
into patchy clouds around the star, causing deep declines  when 
a cloud forms along the line of sight, eclipsing the star (Clayton 1996).  
In addition to massive declines, RCB stars also experience pulsation-induced 
smaller-scale variations ($\sim$0.1 mag) at regular intervals on 
the order of 40-100 days (Lawson \& Kilkenny 1996).

Evidence from polarization observations, geometrical considerations, and 
decline timescales indicates that dust forms close to the photosphere 
(Whitney et al. 1992; Clayton 1996).  Yet effective temperatures of the stars range from 
4,000 to 20,000 K creating an environment thought to be too hot for dust 
grain nucleation to occur.  Thus, a cooling mechanism must exist that enables 
near-photosphere dust condensation.   One suggestion is that low-temperature pockets 
in RCB atmospheres are created by shock-induced non-equilibrium conditions 
resulting from pulsations (Goeres \& Sedlmayr 1992; Woitke, Goeres \& 
Sedlmayr 1996).  In addition, the frequently mentioned 1500 K dust 
condensation temperature is based on solar conditions, where H is abundant.  
Dust condensation temperature increases with the C/H ratio (Donn 1967).
RCB stars have C/H ratios that are 10-50 times greater than the solar 
ratio resulting in a dust condensation temperature that is likely to be 
higher than 1500 K.

It has also been suggested that CO may have a role in triggering dust 
formation (Clayton 1996; Woitke et al. 1996).  CO has a large 
dissociation energy.  
In an environment where 
carbon is significantly more abundant than oxygen, such as in
RCB atmospheres, almost all molecularly-bonded oxygen atoms exist in the 
form of CO, making it the most abundant polar molecule by two orders of 
magnitude (Woitke et al. 1996).  
In high abundance, CO dominates radiative heating and cooling. 
When combined with pulsation-induced
density and temperature changes, an environment suited for dust 
formation may exist in a CO-rich cloud.

Until now, the only published detection of CO in RCB stars was at
UV wavelengths, in the form of the A-X 4th positive system absorption 
bands (Clayton et al. 1999; Clayton \& Ayres 2001).  V854 Cen shows strong UV CO bands, while in RY Sgr, the bands are present but much weaker. 
Clayton et al. (1999) conclude that the weak features seen in the spectrum of RY Sgr are probably characteristic of the normal atmospheric abundance of CO outside of dust formation events.
Unsuccessful searches for IR vibration-rotation bands of CO have been made in RCB stars, including R CrB and RY Sgr (Wilson, Schwartz, \& Epstein 1973; Zuckerman et al. 1978; Knapp et al. 1982; Lambert 1986; Zuckerman \& Dyck 1986; Wannier et al. 1990; Rao et al. 1991). Lambert (1986) reported that the hydrogen-deficient carbon star, HD 182040, shows CO absorption bands in the near-IR, while R CrB, itself, shows no evidence for CO bands. 
The first overtone bands of CO have been detected in the final-helium-flash stars, FG Sge and Sakurai's object (Hinkle, Joyce, \& Smith 1995; Eyres et al. 1998). 
In this paper, we report new observations of the first and second overtone CO bands in the H- and K-bands in RCB stars.

\section{Observations and Reductions}

Near-IR spectra were obtained of eight RCB stars and the RCB-like star, DY Per, between 1998 and 2001. The final-helium-flash star, FG Sge, which is in an RCB-like phase at present was also observed.  The observations of the RCB stars and FG Sge are summarized in Table 1. A majority of the spectra are K-band but also include some J- and H-band spectra. 
Most of the spectra were obtained at  the Steward Observatory's 90-inch Bok telescope at Kitt Peak, Arizona, with FSpec, a cryogenic long-slit near-IR spectrometer utilizing a NICMOS3 256 x 256 array (Williams et al. 1993).  Two gratings were used, 300 and 600 lines mm$^{-1}$, giving resolving powers of $\sim$800 and 3000, respectively. 

A few spectra were obtained using the NIRIS at the Shane
3-m telescope at Lick Observatory (Rudy, Puetter, \& Mazuk 1999).  NIRIS uses two separate channels, divided at 1.38 \micron~by a beam splitter, to provide nearly continuous coverage between 0.8 and 2.5 \micron. Each channel has its own collimator, grating, camera, and HgCdTe detector array. Within each channel the spectral resolution is nearly constant.  A 2.7\arcsec~ slit width was used, resulting in a resolution of 16 \AA~for the blue channel and 37 \AA~for the red channel.  

For both instruments, wavelength calibration was achieved by using features from arcs, telluric absorption features, and OH lines from the night sky. The observations were reduced by dividing the spectrum by that of a nearby solar-type comparison star to remove instrumental response and the effects of atmospheric absorption. Features intrinsic to the comparison star are removed using a solar spectrum
but in any case, the region around 2.3 \micron~has little intrinsic absorption in the standard stars. 
The signal-to-noise ratio in the spectra is typically in the range 30-50. 

Figure 1 shows the NIRIS spectra covering the spectral range 0.8 - 2.5 \micron~for SV Sge and ES Aql.  
The features marked are listed in Table 2. The other stars in the sample have only H- and K-band spectra. They are plotted in Figure 2 with the exception of U Aqr. Its spectrum has a significantly lower S/N than the other stars. SU Tau, XX Cam, and U Aqr  have only K-band spectra. 


\section{Results and Discussion}

The spectra in Figure 1, stretching from 0.8 to 2.5 \micron, show vividly how prevalent carbon is in the spectra of RCB stars. It is well known from visible spectra that the Swan bands of C$_2$, and the violet and red bands of CN are strongest in the coolest ($\sim$5000 K) RCB stars (Clayton 1996; Alcock et al. 2001). SV Sge and ES Aql  both belong to the cool RCB stars and their molecular bands are strong  throughout the spectral region covered. These are the first spectra covering such a wide spectral region from the red to the K-band for any RCB stars. A similar spectrum over the same wavelength region is shown by Sakurai's object, another final flash star.  Its spectrum in 1997 was very similar to an RCB star but it has evolved since (Eyres et al. 1998; Geballe et al. 2002). One important difference in the 1997 spectrum of Sakurai's object is the presence of $^{13}$C which is not seen in any RCB star (Clayton 1996). One of the key observations by which RCB stars can be distinguished from carbon stars is by 
$^{13}$C (Clayton 1996; Alcock et al. 2001). RCB stars show no isotopic $^{13}$C molecular bands. 
Carbon stars show $^{13}$CO bands as well as $^{12}$CO (e.g., Lan\c{c}on \& Wood 2000; F\"{o}rster Schreiber 2000).

Higher resolution spectra of the H- and K-band spectral regions are shown in Figure 2 for six RCB stars, as well as for FG Sge and DY Per.  In the K-band, the warm ($>$6500 K)  RCB stars, R CrB, RY Sgr, SU Tau and XX Cam,  show no sign of CO absorption, while the cool ($<$6000 K) RCB stars, SV Sge, ES Aql, and Z UMi,  show strong absorption bands.  U Aqr, not shown in Figure 2, may have weak CO bands but the low S/N of its K-band spectrum prevents a definitive detection. The final-flash star, FG Sge, which has a T$_{eff}\sim$5500 K (Kipper \& Klochkova 2001), has significant CO bands, similar to the cool RCB stars.  In the H-band, the data are noisier and sample fewer stars but the same behavior is seen.  The cool stars, SV Sge and Z UMi show significant CO bands while the warm stars, R CrB and RY Sgr, do not. In this region, DY Per also shows CO bands but FG Sge does not. 

The detection of CO in the RCB stars provides yet another similarity of FG Sge to this class of variable stars (Gonzalez et al. 1998). The CO bands are weaker in FG Sge than in the RCB stars with similar T$_{eff}$. There is no sign of $^{13}$CO absorption but the $^{12}$CO bands are relatively weak and the $^{13}$CO bands would be weaker. 
DY Per represents a cool, slow-declining subclass of the RCB stars (Alcock et al. 2001; Keenan \& Barnbaum 1997). It  also has strong CO absorption bands. DY Per shows $^{13}$C in the Swan bands in the visible (Alcock et al. 2001) but there is no significant $^{13}$CO absorption in the spectra presented here. It is not clear whether the DY Per and RCB stars are related. 
One hydrogen-deficent carbon star, HD 137613, shows strong isotopic bands of $^{12}$C$^{18}$O as well as $^{12}$C$^{16}$O (Clayton et al. 2005). None of the stars in this study show similar strong isotopic bands although weak bands could be present. 

As mentioned above, CO could have a pivotal role in the formation of dust around RCB stars (Clayton 1996; Woitke et al. 1996). However, the detection of CO absorption in RCB stars is not necessarily an  indication of dust formation.  CO is present in the atmospheres of cool stars, including the Sun (e.g., Ayres \& Rabin 1996).  
The CO absorption bands can been seen in 
 flux distributions constructed
from model atmosphere calculations (Asplund et al. 1997, 2000). The models have the usual assumptions, i.e., constant flux, plane parallel atmosphere, in hydrostatic and local thermodynamic equilibrium, and use the same RCB star abundances as Asplund et al. (1997).  
The model is a new version of the MARCS model atmosphere code which
has full opacity sampling and line-blanketing, representing
an improvement over that presented in Asplund et al. (1997).  
For comparison with our spectra, we calculated a grid of models from T$_{eff}$=5000 to 7500 for both $log~g$=0.5 and 1.5. 
There is very little difference seen in the strength of the CO absorption bands at a given T$_{eff}$  between $log~g$ = 0.5 and 1.5. 
The models presented here are
not spectrum synthesis calculations nor do they use individual abundances of the sample stars. We merely computed the
opacity sampling fluxes from the models using the standard RCB star abundances for a grid of T$_{eff}$ (Asplund et al. 1997). 
We compare examples of these model flux distributions to observed spectra in Figure 3.
It can be seen that there is good qualitative agreement between the model and the observed spectra. In the models, strong CO bands are seen at 5000 K and progressively weaken as T$_{eff}$ increases. No CO absorption is seen in models with T$_{eff}$ = 6500, 7000 and 7500 K. Similarly, our sample RCB stars with T$_{eff}>$6500 K show no sign of CO bands. 
We plan to obtain higher S/N spectra with better spectral resolution of the RCB stars that can be used for synthesis modeling. 

Variations in the strength of the CO bands correlated with the decline activity of an RCB star would be evidence of the role of CO in dust formation. Six stars in our sample were observed at least twice in the K-band.  SV Sge has the best temporal coverage having K-band spectra obtained in 1998, 1999, 2000 and 2001. The sample stars are all monitored by the AAVSO allowing us to correlate the strengths of CO bands with the dust formation activity of the stars.  The observations were not timed to the decline activity of the sample stars but they include stars at both maximum light and in deep declines. For XX Cam  and RY Sgr, both spectra were obtained at maximum light.  XX Cam is the most inactive RCB star, not having had a decline for many years (Clayton 1996). RY Sgr had a deep decline about a year after the CO observations.  SU Tau was observed during a period of high activity. The first spectrum was obtained as it was recovering from one deep decline, and the second as it was in a subsequent deep decline.  R CrB was also very active during the observation period. It was first observed while in a 4-magnitude decline and then at maximum light between two declines. DY Per was observed three times, all during declines.  SV Sge was observed first during a recovery from a deep decline and then three times at maximum light. The last spectrum was obtained about 50 days before the onset of a deep decline.  We measured the equivalent widths of the first overtone bands for these stars at each of the observed epochs.  None of the stars showed variations greater than 20\% in their CO absorption band strengths. Variations of that size are within the measurement uncertainty of the spectra. 

These results are not definitive since the sampling was extremely sparse for any individual star. Nevertheless, no significant changes in the CO band strengths were seen in any sample star whether at maximum light or in a decline. 
The CO band strengths seen in the sample stars are consistent with the predictions from the atmosphere models at the corresponding T$_{eff}$.
It should be noted that large changes in the strengths of C$_2$ and CN bands in the visible, do show significant changes due to temperature changes caused by pulsations in the RCB star atmospheres (Clayton et al. 1995). No such variations are seen in the sparse data presented here for the near-IR CO bands.

\section{Conclusions}

A survey for CO  in near-IR spectra of a small sample of RCB stars has been completed.\\
\noindent
 $\bullet$We report the first detection in RCB stars of the CO first and second overtone bands in H- and K-band spectra.  Strong CO absorption bands are seen in three cool (4000-5000 K) RCB stars while no bands were  detected in four warm (6500-7250 K) RCB stars.\\
\noindent
$\bullet$The strengths of the CO absorption bands are strongly correlated with  T$_{eff}$ corresponding to the predictions of line-blanketed model atmospheres for the RCB stars.\\
\noindent
$\bullet$No temporal variations in the strength of the CO bands are seen during dust formation episodes in individual stars. 

\acknowledgements

This project was supported by the NSF/REU grant AST-0097694 and the 
Nantucket Maria Mitchell Association. We would also like to thank Vladimir 
Strelnitski, Director, Maria Mitchell Observatory.
We acknowledge with thanks the variable star observations  from the AAVSO International Database contributed by  observers worldwide and used in this research.
Partial support was provided by the Independent Research and Development program of The Aerospace Corporation. We thank the referee for several useful suggestions.

\clearpage

\begin{figure*}
\figurenum{1}
\epsscale{1.0}
\vspace{1.cm}
\plotone{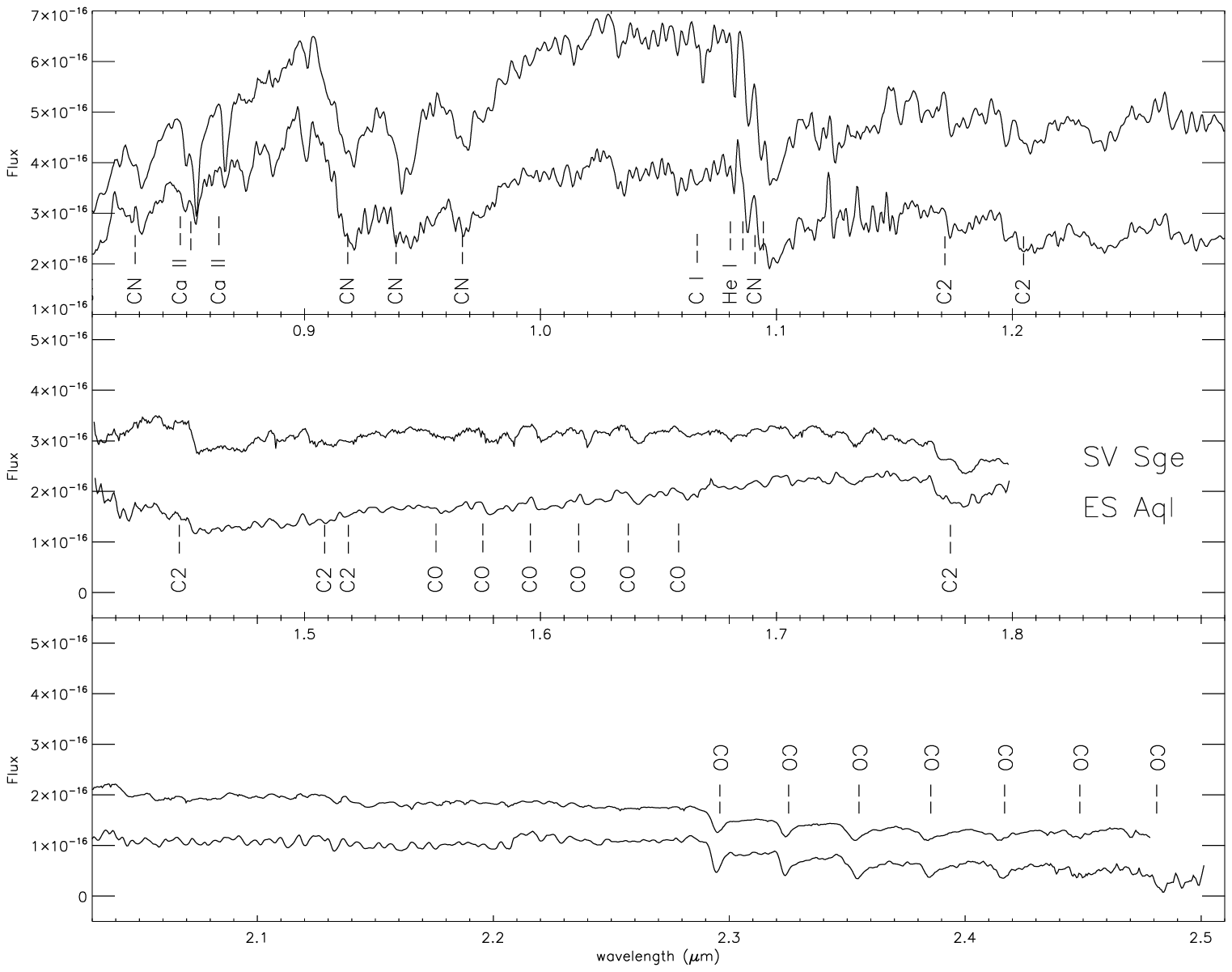}
\caption{NIRIS spectra of SV Sge and ES Aql. The fluxes are in $erg ~cm^{-2}~s^{-1}$ \AA$^{-1}$. The fluxes for ES Aql have been multipied by a factor of 8. In addition, the spectrum of ES Aql in the bottom panel has been shifted downward by 1.0 x 10$^{-16}$ $erg ~cm^{-2}~s^{-1}$ \AA$^{-1}$ for clarity. The features marked are listed in Table 2. }
\end{figure*}

\begin{figure*}
\figurenum{2}
\epsscale{1.0}
\vspace{1.cm}
\plottwo{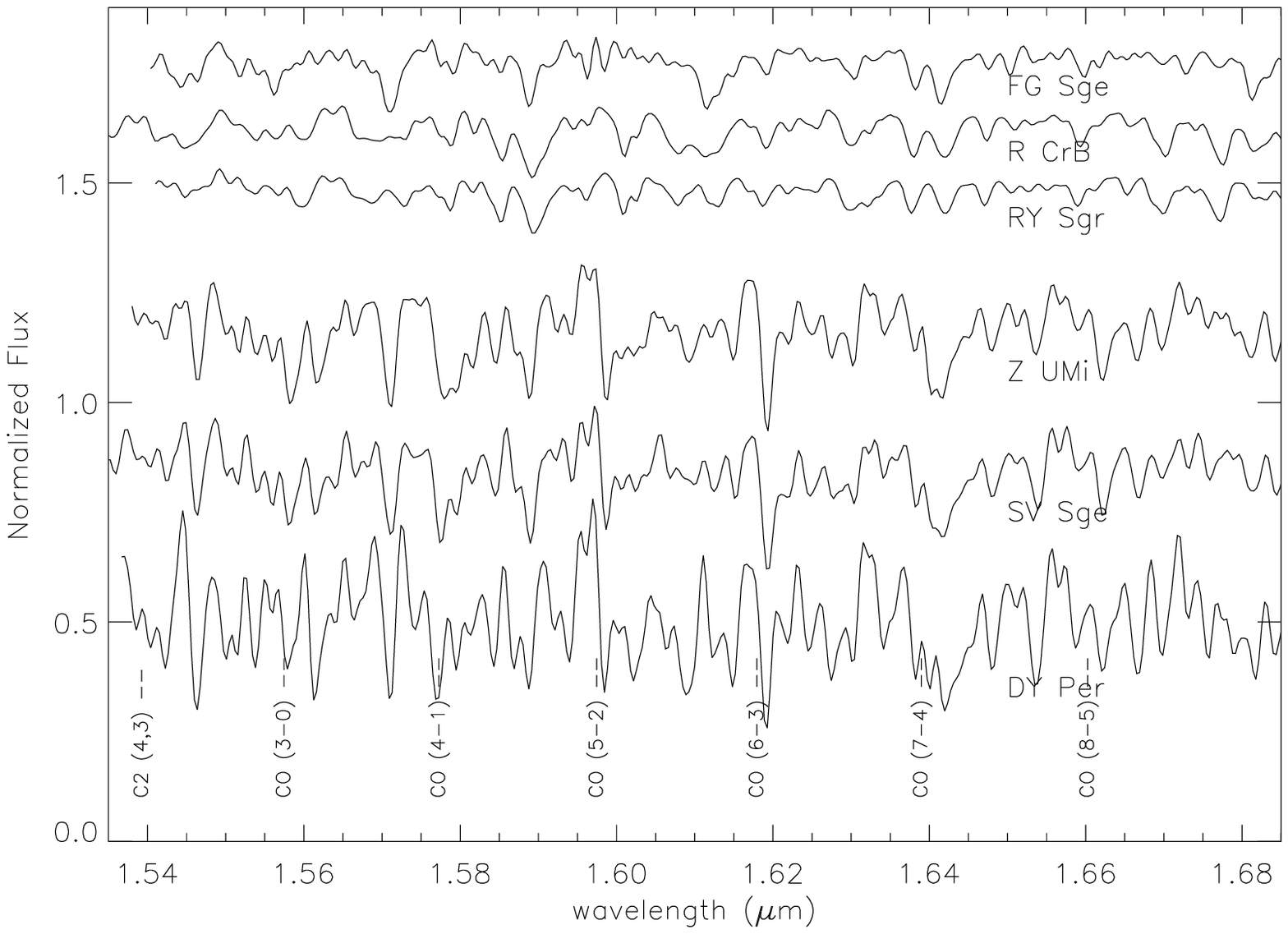}{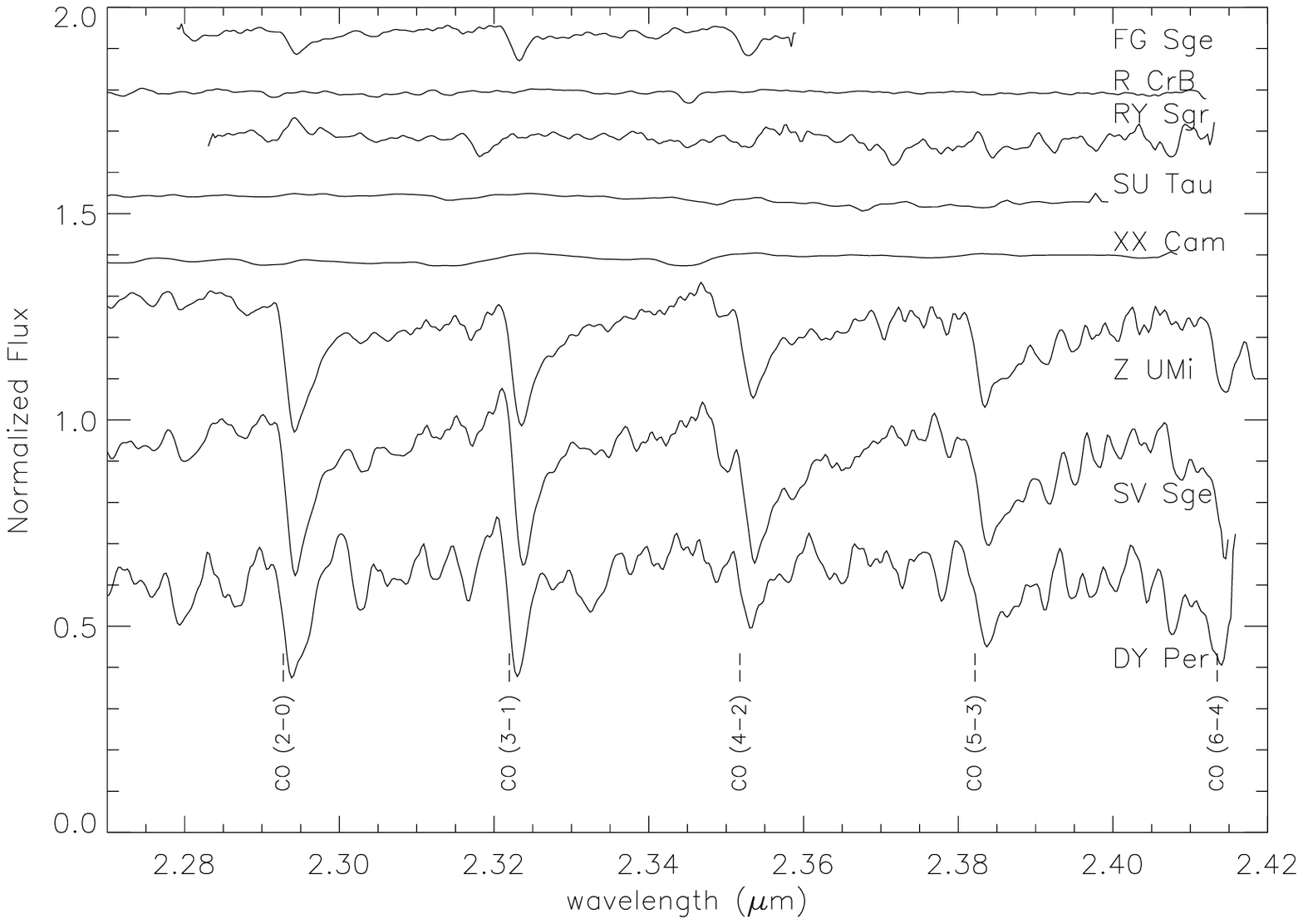}
\caption{H-band (left panel) and K-band (right panel) spectra of a sample of RCB stars. The spectra
have been shifted up or down for clarity.}
\end{figure*}

\begin{figure*}
\figurenum{3}
\epsscale{1.0}
\vspace{1.cm}
\plotone{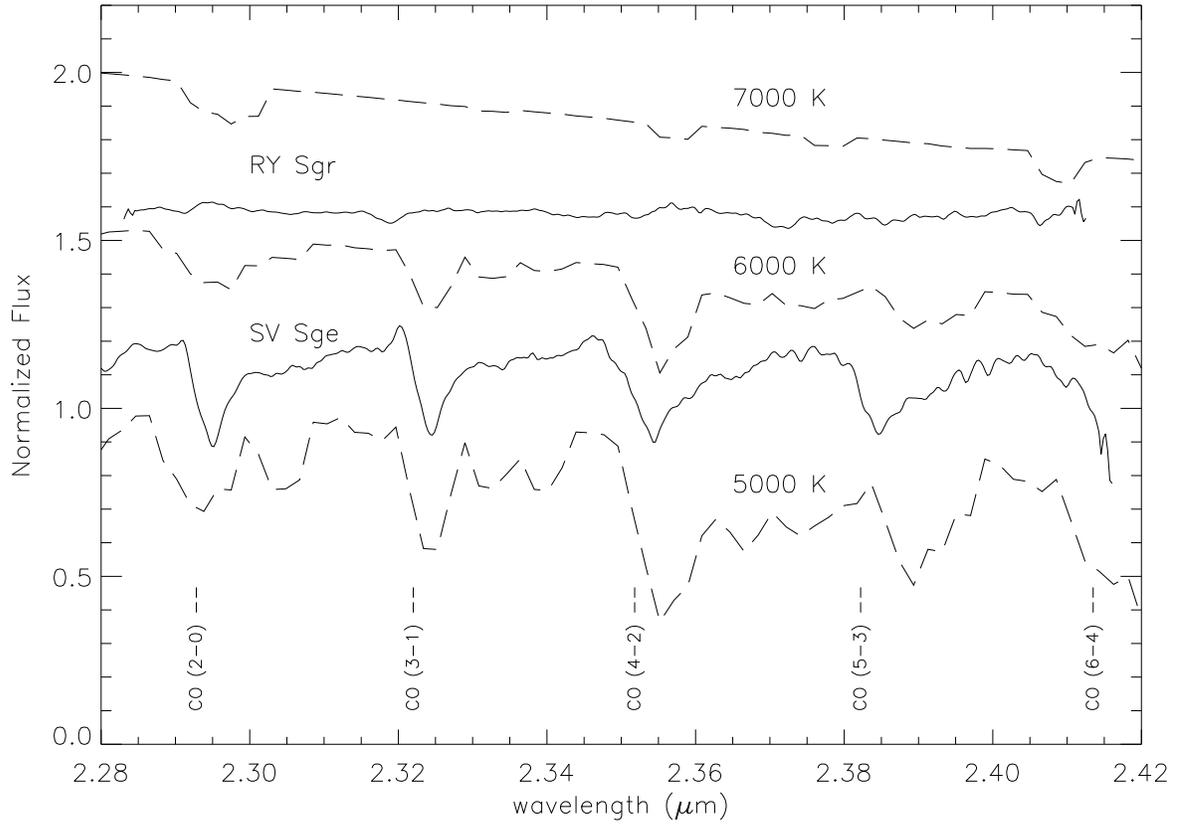}
\caption{K-band spectra of RY Sgr and SV Sge (solid lines) plotted with three line-blanketed atmospheric models (dashed lines) for different effective temperatures (Asplund et al. 1997). The observed spectra have been smoothed to the resolution of the models. See text.}
\end{figure*}

\clearpage
\oddsidemargin=-1cm
\tabletypesize{\scriptsize}

\begin{deluxetable}{cccccc}




\tablecaption{Observations}

\tablenum{1}

\tablehead{\colhead{Star}&\colhead{T$_{eff}$\tablenotemark{a} (K)} & \colhead{UT Date} & \colhead{Instr.} & 
\colhead{Band}
& \colhead{Grating}}
\startdata
DY Per &4700&10/11/98	&	FSpec	&	K	&	300 \\
	&&6/20/00	&	FSpec	&	HK	&	600 \\
	&&2/7/99	&	FSpec	&	K	&	300 \\
ES Aql &$\sim$5000 	&	7/9/01	&	NIRIS	&	JHK	&	\nodata \\
FG Sge &5500 	&	6/19/00	&	FSpec	&	H	&	600 \\
R CrB &6750 	&	2/7/99	&	FSpec	&	K	&	300 \\
	&&	6/21/00	&	FSpec	&	HK	&	600 \\
	&&	4/1/99	&	FSpec	&	H	&	300 \\
RY Sgr &7250 	&	10/10/98	&	FSpec	&	K	&	300 \\
	&&	6/18/00	&	FSpec	&	HK	&	600 \\
SU Tau &6500 	&	10/11/98	&	FSpec	&	K	&	300 \\
	&&	2/7/99	&	FSpec	&	K	&	300 \\
SV Sge &4000 &	10/10/98	&	FSpec	&	K	&	300 \\
	&&	6/19/00	&	FSpec	&	H	&	600 \\
	&&	6/21/00	&	FSpec	&	K	&	600 \\
	&&	9/99	&	NIRIS	&	JHK	&	\nodata \\
	&&	7/9/01		&NIRIS	&	JHK	&	\nodata \\
U Aqr& 6000	&	6/20/00	&	FSpec	&	K	&	600 \\
XX Cam &6600 	&	10/11/98	&	FSpec	&	K	&	300 \\
	&&	2/6/99		&FSpec	&	K	&	300 \\
Z Umi &$\sim$5000 		&	6/22/00	&	FSpec	&	H	&	600 \\
	&&	6/18/00		&FSpec	&	K	&	600 \\
\enddata
\tablenotetext{a}{Estimates of T$_{eff}$ were obtained from Asplund et al. (1997, 2000),  Bergeat, Knapik, \& Rutily (2001), Keenan \& Barnbaum (1997) and Kipper \& Klochkova (2001). ES Aql and Z Umi have not been modeled.)}



\end{deluxetable}

\clearpage

\begin{deluxetable}{ll}




\tablecaption{Atomic \& Molecular features Identified}

\tablenum{2}

\tablehead{\colhead{Species} & \colhead{wavelength (\micron)}}
\startdata
CN Red +2   &  0.8102\\
CN Red +2     &  0.8307\\
Ca II  &  0.8498\\
Ca II  &  0.8543\\
Ca II &   0.8662\\
CN Red +1    &   0.9208\\
CN Red +1    &   0.9413\\
CN Red +1    &   0.9694\\
C I     & 1.0689\\
He I  &   1.0830\\
CN Red 0    &   1.0883\\
CN Red 0    &   1.0934\\
CN Red 0    &   1.0969\\
C2 Balik-Ramsay  +2 &    1.1739\\
C2 Phillips 0  &    1.2072\\
C2  Balik-Ramsay +1 &1.4081\\
C2  Balik-Ramsay +1 &1.4494\\
C2 Phillips  -1    &  1.5040\\
C2 Phillips  -1   &  1.5110\\
C2 Phillips  -1  &  1.5210\\
CO (3-0) &1.5582\\
CO (4-1)& 1.5780\\
CO (5-2)& 1.5982\\
CO (6-3) &1.6187\\
CO (7-4) &1.6397\\
CO (8-5) &1.6610\\
C2 Balik-Ramsay 0   &   1.7790\\
CO (2-0)& 2.2935\\
CO (3-1)& 2.3227\\
CO (4-2)& 2.3535\\
CO (5-3)& 2.3829\\
CO (6-4)& 2.4142\\
CO (7-5) &2.4461\\
CO (8-6) &2.4787\\
\enddata



\end{deluxetable}

\end{document}